\documentclass[aps,prc,twocolumn,groupedaddress,floatfix,10pt]{revtex4-1}

\usepackage{graphicx}
\usepackage{dcolumn}
\usepackage{bm}
\usepackage{morefloats}
\usepackage{multirow}
\usepackage{amsfonts}
\usepackage{amssymb}
\usepackage{amsmath}
\usepackage{bbold}
\usepackage{fix-cm}
\usepackage{mathptmx} 
\usepackage[T1]{fontenc}
\usepackage[colorlinks,allcolors=blue]{hyperref}

\begin{document}

\title{Quantal Nucleon Diffusion I: Central Collisions of Symmetric Nuclei}
\author{S. Ayik$^{1}$}\email{ayik@tntech.edu}
\author{O. Yilmaz$^{2}$}
\author{B. Yilmaz$^{3}$}
\author{A. S. Umar$^{4}$}
\affiliation{
$^{1}$Physics Department, Tennessee Technological University, Cookeville, TN 38505, USA \\
$^{2}$Physics Department, Middle East Technical University, 06800 Ankara, Turkey \\
$^{3}$Physics Department, Faculty of Sciences, Ankara University, 06100 Ankara, Turkey \\
$^{4}$Department of Physics and Astronomy, Vanderbilt University, Nashville, Tennessee, 37235, USA}

\date{\today}

\begin{abstract}
Quantal diffusion mechanism of nucleon exchange is studied in the central collisions of several
symmetric heavy-ion collisions in the framework of the Stochastic Mean-Field (SMF) approach. Since at
bombarding energies below the fusion barrier, di-nuclear structure is maintained, it is possible to
describe nucleon exchange as a diffusion process familiar from deep-inelastic collisions. Quantal
diffusion coefficients, including memory effects, for proton and neutron exchanges are extracted
microscopically employing the SMF approach. The quantal calculations of
neutron and proton variances are compared with the semi-classical results.
\end{abstract}


\maketitle

\section{Introduction}
Multi-nucleon exchange is an important mechanism in deep-inelastic heavy-ion collisions. Considerable effort has been spent for description of
deep-inelastic collisions in terms of nucleon transport models\,\cite{ayik1976,randrup1979,merchant1982,schroder1981}. More recently, it has been realized that
multi-nucleon exchange is an important process in the quasi-fission reactions of heavy-ions\,\cite{loveland2007}. For heavy systems, at collision energies near fusion
barrier, the compound nucleus formation is severely inhibited by the quasi-fission mechanism. The colliding ions stick together for a long time,
but separate without going through the compound nucleus formation. During the long contact times many nucleons are exchanged between the projectile and
the target nuclei. In multi-nucleon exchange reactions, it is possible to study charge equilibration driven by the nuclear symmetry-energy\,\cite{rizzo2011}, and to produce
very neutron rich heavy-ions\,\cite{sahm1984,schmidt1991}. A number of models have been developed for the description of the reaction mechanism in the multi-nucleon
transfer process in quasi-fission reactions\,\cite{adamian2003,zagrebaev2007,aritomo2009,diaz-torres2001}. The mean-field approach and the time-dependent Hartree-Fock (TDHF) theory
provide a microscopic basis for describing heavy-ion reaction mechanism at low bombarding
energies\,\cite{negele1982,simenel2012}.
Within the last few years the TDHF approach has been utilized for studying the dynamics of
quasifission~\cite{wakhle2014,oberacker2014,umar2015a,umar2015c}.
Particularly, the study of quasifission is showing a great promise to provide
insight based on very favorable comparisons with experimental data.
However, in the mean-field approximation the collective aspects of collision dynamics is treated semi-classically, and fluctuations of the macroscopic variables are severely inhibited.
To remedy this problem one must go beyond TDHF~\cite{tohyama2002a,simenel2011,lacroix2014}.

In the recently developed Stochastic Mean-Field (SMF) approach, dynamical description is extended beyond the mean-field approximation by incorporating the
initial fluctuations\,\cite{ayik2008,lacroix2014}. In a number of studies, it has been demonstrated that the SMF approach alleviates the drawbacks of the standard
mean-field approach and improves the description of nuclear collisions dynamics including fluctuation mechanism of the collective motion
\cite{ayik2009,washiyama2009b,yilmaz2011,yilmaz2014,ayik2015a,lacroix2012,lacroix2014}. Most applications have been carried out for collisions where a di-nuclear
structure is maintained. In this case it is possible to define macroscopic variables by a geometric projection procedure with the help of the window
dynamics. The SMF approach gives rise to a Langevin description for the evolution of macroscopic variables\,\cite{gardiner1991,weiss1999}. In most analysis of
the nucleon diffusion mechanism, the deduced Langevin description has been applied by calculating transport coefficient in the semi-classical
approximation and neglecting the memory effects. In a recent work, we investigated nucleon exchange mechanism for the central collisions of several
symmetric systems in the quantal framework of the SMF approach under certain approximation\,\cite{ayik2015}. In this work, we consider central
collisions of the symmetric systems below the fusion barrier as well, but improve the quantal description of the diffusion mechanism to a large
extend. We extract quantal diffusion coefficients for proton and neutron transfers including memory effects from the SMF approach. In symmetric
collisions, the mean values of the proton and neutron numbers of the outgoing fragments do not change. However, as result of nucleon exchange,
outgoing fragments exhibit charge and mass distributions around their initial values. We carry out calculations for the variance of neutron and proton
distributions of the outgoing fragments in the central collisions of ${}^{28}$O + $^{28}$O, $^{40}$Ca + $^{40}$Ca, $^{48}$Ca + $^{48}$Ca, and
$^{56}$Ni + $^{56}$Ni systems at bombarding energies slightly below their fusion barriers, and compare the results with the corresponding semi-classical calculations.

In Sec.\,\ref{sec:2}, we present a brief description of the quantal nucleon diffusion mechanism based on the SMF approach. In Sec.\,\ref{sec:3}, we present derivation
of quantal expression for proton and neutron diffusion coefficients. The result of calculations is reported in Sec.\,\ref{sec:4}, and conclusions are given in
Sec.\,\ref{sec:conc}.

\section{Diffusion Mechanism}
\label{sec:2}
In heavy-ion collisions when the system maintains a binary structure, the reaction evolves primarily by nucleon exchange through the window between the
projectile-like and target-like fragments. It is possible to analyze the nucleon exchange mechanism by employing nucleon diffusion concept based on the
SMF approach. In the SMF approach, the standard mean-field description is extended by incorporating the mean-field fluctuations in terms of generating
an ensemble of events according to quantal and thermal fluctuations in the initial state. Instead of following a single event specified by fixed
initial conditions, in the SMF approach an ensemble of events are propagated specified in terms of the quantal and thermal fluctuations of the initial
state, for details please refer to\,\cite{ayik2008,ayik2009,washiyama2009b,yilmaz2011,yilmaz2014,ayik2015a}.  In extracting transport coefficients for nucleon
exchange, we take the proton and neutron numbers of projectile-like fragments $Z_{1}^{\lambda } ,N_{1}^{\lambda } $ as independent variables, where
$\lambda $ indicates the event label . We can define the proton and neutron numbers of the projectile-like fragments in each event by integrating over
the nucleon density on the projectile side of the window.  In the central collisions of symmetric systems, the window is perpendicular to the
collision direction taken as the $x$-axis and the position of the window is fixed at the origin of the center of mass frame at $x_{0} =0$.
The proton and neutron numbers of the projectile-like fragments are defined as,
\begin{eqnarray} \label{eq1}
\left(\begin{array}{c} {Z_{1}^{\lambda } (t)} \\ {N_{1}^{\lambda } (t)} \end{array}\right)=\int d^3r \theta (x-x_{0} ) \left(\begin{array}{c} {\rho _{p}^{\lambda } (\vec{r},t)} \\ {\rho _{n}^{\lambda } (\vec{r},t)} \end{array}\right).
\end{eqnarray}
Here, $\rho _{p}^{\lambda } (\vec{r},t)$ and $\rho _{n}^{\lambda } (\vec{r},t)$ are the local densities of protons and neutrons, and $x_{0} =0$.
According to the SMF approach, the proton and neutron numbers of the projectile-like fragment follow a stochastic evolution according to the Langevin
equations,
\begin{eqnarray} \label{eq2}
\frac{d}{dt} \left(\begin{array}{c} {Z_{1}^{\lambda } (t)} \\ {N_{1}^{\lambda } (t)} \end{array}\right)&=&\int d^{3} rg(x)\left(\begin{array}{c} {J_{x,p}^{\lambda } (\vec{r},t)} \\ {J_{x,n}^{\lambda } (\vec{r},t)} \end{array}\right)\nonumber\\
&=&\left(\begin{array}{c} {v_{p}^{\lambda } (t)} \\ {v_{n}^{\lambda } (t)} \end{array}\right).
\end{eqnarray}
In this expression, we introduce a smoothing function $g(x)$ for convenience,
\begin{eqnarray} \label{eq3}
g(x)=\frac{1}{\sqrt{2\pi}\kappa}\exp\left(-\frac{x^2}{2\kappa^2} \right).
\end{eqnarray}
In the limit $\kappa \rightarrow 0$, $g(x)$ becomes a delta function $g(x)\rightarrow \delta(x)$.  The right hand side of Eq.\,(\ref{eq2}) denotes the
proton, $v_{p}^{\lambda } (t)$, and neutron, $v_{n}^{\lambda } (t)$, drift coefficients for the event $\lambda $, which are determined by the proton and
the neutron current densities, $J_{x,p}^{\lambda } (\vec{r},t)$, $J_{x,n}^{\lambda } (\vec{r},t)$, through the window for that event.  In the SMF
approach, the fluctuating proton and neutron current densities are defined as,
\begin{eqnarray} \label{eq4}
J_{x,\alpha }^{\lambda } (\vec{r},t)=\frac{\hbar }{m} \sum _{ij\in \alpha }\text{Im}\left[\Phi _{j}^{*} (\vec{r},t;\lambda )\nabla _{x} \Phi _{i} (\vec{r},t;\lambda )\rho _{ji}^{\lambda }\right] .
\end{eqnarray}
Here, and in the rest of the paper, we use the label $\alpha =p,n$ for the proton and neutron states. The parameter $\kappa $ of the Gaussian smoothing
function is determined by setting typical particle-hole matrix elements of proton and neutron currents through the window
\begin{eqnarray} \label{eq5}
\Lambda _{ji}^{\lambda } (t)=\int d^{3} rg(x)\; \text{Im}\left[\Phi _{j}^{*} (\vec{r},t;\lambda )\nabla _{x} \Phi _{i} (\vec{r},t;\lambda )\right]
\end{eqnarray}
to their values obtained at $\kappa \to 0$ as $g(x)\to \delta (x)$. It turns out that this limiting value equals to smoothed value by means of a
Gaussian with a dispersion given by value $\kappa =1.0$\,fm. This value is in the order of lattice spacing, which indicates the numerical calculations
implicitly involve such a smoothing mechanism. Drift coefficients fluctuate from event to event due to stochastic elements of the initial density
matrix $\rho _{ji}^{\lambda } $ and also due to the different sets of the wave functions in different events.  As a result, there are two sources for
fluctuations of the nucleon current: (i) fluctuations arising from the state dependence of the drift coefficients, which may be approximately
represented in terms of fluctuations of proton and neutron numbers of the di-nuclear system, and (ii) the explicit fluctuations $\delta v_{p}^{\lambda} (t)$ and $\delta v_{n}^{\lambda } (t)$,
which arise from the stochastic part of the proton and neutron currents\,\cite{ayik2015a,simenel2011}. In the present
work, we investigate the nucleon diffusion mechanism for the central collisions of light heavy-ions. Due to the relatively short collision times,
fluctuations driven by the symmetry energy are small. Therefore, we neglect the fluctuation mechanism due to the state dependence of the
drift coefficients and include only the explicit fluctuations arising from the stochastic part of the current densities. Equations for the mean values
of proton, $Z_{1} (t)=\overline{Z_{1}^{\lambda } (t)}$, and neutron, $N_{1} (t)=\overline{N_{1}^{\lambda } (t)}$, numbers of the projectile-like fragments
are obtained by taking the ensemble averaging of the Langevin Eq.\,(\ref{eq2}). Here and below, the bar over a quantity indicates the average over the
generated ensemble. For small amplitude fluctuations, we obtain the usual mean-field result given by the TDHF equations,
\begin{eqnarray} \label{eq6}
\frac{d}{dt} \left(\begin{array}{c} {Z_{1} (t)} \\ {N_{1} (t)} \end{array}\right)&=&\int d^{3} rg(x)\left(\begin{array}{c} {J_{x,p} (\vec{r},t)} \\ {J_{x,n} (\vec{r},t)} \end{array}\right)\nonumber\\
&=&\left(\begin{array}{c} {v_{p} (t)} \\ {v_{n} (t)} \end{array}\right).
\end{eqnarray}
Mean values of the current densities of protons and neutrons along the collision direction are given by,
\begin{eqnarray} \label{eq7}
J_{x,\alpha } (\vec{r},t)=\frac{\hbar }{m} \sum _{h\in \alpha }\text{Im}\left[\Phi _{h}^{*} (\vec{r},t)\nabla _{x} \Phi _{h} (\vec{r},t)\right],
\end{eqnarray}
where the summation $h$ runs over the occupied states originating both from the projectile and the target nuclei. The drift coefficients   $v_{p} (t)$
and $v_{n} (t)$ denote the net proton and neutron currents across the window, respectively.  In order to calculate the fluctuations of the proton and neutron numbers
of the fragments we linearize the Langevin Eq.\,(\ref{eq2}) around the mean values $v_{p} (t)$ and $v_{n} (t)$, and keep only the stochastic part of
the currents to obtain,
\begin{eqnarray} \label{eq8}
\frac{d}{dt} \left(\begin{array}{c} {\delta Z_{1}^{\lambda } (t)} \\ {\delta N_{1}^{\lambda } (t)} \end{array}\right)=\left(\begin{array}{c} {\delta v_{p}^{\lambda } (t)} \\ {\delta v_{n}^{\lambda } (t)} \end{array}\right).
\end{eqnarray}
The variances of neutron and proton distribution of projectile fragments are defined as $\sigma _{nn}^{2} (t)=\overline{\left(N_{1}^{\lambda } -N_{1}
\right)^{2} }$ and $\sigma _{pp}^{2} (t)=\overline{\left(Z_{1}^{\lambda } -Z_{1} \right)^{2} }$.  Multiplying both side of
Eq.\,(\ref{eq8}) by
$N_{1}^{\lambda } -N_{1} $ and $Z_{1}^{\lambda } -Z_{1} $, and taking the ensemble averages, we find the proton and neutron variances are determined by
\begin{eqnarray} \label{eq9}
\frac{d}{dt} \sigma _{\alpha \alpha }^{2} (t)=2D_{\alpha \alpha } (t),
\end{eqnarray}
where $D_{\alpha \alpha } (t)$ denote the diffusion coefficients of proton and neutron exchanges.

\section{Quantal Diffusion Coefficients for Nucleon Exchange}
\label{sec:3}
The quantal expressions of the proton and neutron diffusion coefficients are determined by the correlation function of the stochastic part of the
drift coefficients according to
\cite{gardiner1991,weiss1999},
\begin{equation} \label{eq10}
D_{\alpha \alpha } (t)=\int _{0}^{t}dt' \overline{\delta v_{\alpha }^{\lambda } (t)\delta v_{\alpha }^{\lambda } (t')}.
\end{equation}
From Eq.\,(\ref{eq4}), the stochastic parts of the drift coefficients are given by,
\begin{equation} \label{eq11}
\delta v_{\alpha }^{\lambda } (t)=\frac{\hbar }{m} \sum _{ij\in \alpha }\int d^{3} r g(x)
\mathrm{Im}\left[\Phi _{j}^{*} (\vec{r},t)\nabla _{x} \Phi _{i} (\vec{r},t)\delta \rho _{ji}^{\lambda }\right].
\end{equation}
In determining the stochastic parts of the drift coefficients, we impose a physical constraint on the summations over single-particle sates. The
transitions among single particle states originating from the projectile or target nuclei do not contribute to the nucleon exchange mechanism. Therefore in Eq.\,(\ref{eq11}),
we restrict the summation as follows: when the summation $i\in T$  runs over the states originating from target nucleus, the summation
$j\in P$ runs over the states originating from the projectile, and vice versa. The main postulate of the SMF approach is that the
 stochastic part of the
elements of the initial density matrix $\delta \rho _{ji}^{\lambda } $ have uncorrelated Gaussian distributions with zero mean values and second
moments determined by\,\cite{ayik2008},
\begin{eqnarray} \label{eq12}
\overline{\delta \rho _{ji}^{\lambda } \delta \rho _{i'j'}^{\lambda } }=\frac{1}{2} \delta _{ii'} \delta _{jj'} \left[n_{i} (1-n_{j} )+n_{j} (1-n_{i} )\right],
\end{eqnarray}
where $n_{j} $ are the average occupation numbers of the single-particle states. Using this result, we can calculate the correlation functions of the
stochastic part of the drift coefficients.  At zero temperature, since the average occupation factor are zero or one, we find that the correlation
functions are given by,
\begin{eqnarray} \label{eq13}
\overline{\delta v_{\alpha }^{\lambda } (t)\delta v_{\alpha }^{\lambda } (t')}&=&\text{Re}\left[\sum _{p\in P,h\in T}A_{ph}^{\alpha } (t)A_{ph}^{*\alpha } (t')\right.\nonumber\\
&&\qquad\left.+\sum _{p\in T,h\in P}A_{ph}^{\alpha } (t)A_{ph}^{*\alpha } (t')  \right].
\end{eqnarray}
We note that, because of orthogonality, the particle states $p$ and the hole states $h$ must carry the same spin and the isospin labels.  The
summation runs over the complete set of the particle and hole states of protons and neutrons. The matrix $A_{ph}^{\alpha } (t)$  is determined from
the particle-hole states of the mean-field Hamiltonian,
\begin{eqnarray} \label{eq14}
A_{ph}^{\alpha } (t)&=&\frac{\hbar }{2m} \int d^{3} r g(x)\left[\Phi _{p}^{*\alpha } (\vec{r},t)\nabla _{x} \Phi _{h}^{\alpha } (\vec{r},t)\right.\nonumber\\
&&\qquad\qquad\qquad\quad\left.-\Phi _{h}^{\alpha } (\vec{r},t)\nabla _{x} \Phi _{p}^{*\alpha } (\vec{r},t)\right].
\end{eqnarray}
With the help of partial integration we can express this matrix element as,
\begin{eqnarray} \label{eq15}
A_{ph}^{\alpha } (t)&=&\frac{\hbar }{m} \int d^{3} r \Phi _{p}^{*\alpha } (\vec{r},t)g(x)
\left[\frac{}{}\nabla _{x} \Phi _{h}^{\alpha } (\vec{r},t)\right.\nonumber\\
&&\qquad\qquad\qquad\qquad\qquad\left.-\frac{x}{2\kappa ^{2} } \Phi _{h}^{\alpha } (\vec{r},t)\right].
\end{eqnarray}
In order to calculate the correlation function Eq.\,(\ref{eq13}) directly, in addition the occupied hole states, we need to evolve a complete set of
particle states.  This is a very difficult task to accomplish. In a previous work, we carried out an approximate description of the correlation function
by calculating it with a set of particle-hole states and increasing the volume of the particle-hole space step by step\,\cite{ayik2015}.  We observed
that the convergence of such a calculation was very slow and required ever increasing computational time to proceed.  Even the results obtained with
sufficiently large particle-hole spaces did not compare favorably with the results of the semi-classical calculations.

Here, we introduce a different approach to evaluate the correlation function Eq.\,(\ref{eq13}).  In the first term of the right
hand side of Eq.\,(\ref{eq13}), we add and subtract the hole contributions to give,
\begin{eqnarray} \label{eq16}
\sum _{p\in P,h\in T}A_{ph}^{\alpha } (t)A_{ph}^{*\alpha } (t')&=&\sum _{a\in P,h\in T}\!\!\!\!A_{ah}^{\alpha } (t)A_{ah}^{*\alpha } (t')\nonumber\\
&&-\!\!\sum _{h'\in P,h\in T}\!\!\!\!\!\!A_{h'h}^{\alpha } (t)A_{h'h}^{*\alpha } (t').
\end{eqnarray}
Here, the summation $a$ is over the complete set of states originating from the projectile. In the first term, we cannot use the closure relation to
eliminate the complete set of single-particle states, because the wave functions are evaluated at different times. However, we note that the
time-dependent single-particle wave functions during short time intervals exhibit nearly a diabatic behavior\,\cite{norenberg1981}. In another way of
stating, that during short time intervals the nodal structure of time-dependent wave functions do not change appreciably. Most dramatic diabatic
behavior of the time-dependent wave-functions is apparent in the fission dynamics. The Hartree-Fock solutions force the system to follow the diabatic
path, which prevents the system to break up into fragments.  As a result of these observations, during the short time interval $\tau =t-t'$ evolutions,
in the order of the correlation time, a diabatic approximation for the time dependent wave-functions can be done by shifting the time backward (or forward)
according to,
\begin{eqnarray} \label{eq17}
\Phi _{a} (\vec{r},t')\approx \Phi _{a} (\vec{r}-\vec{u}\tau ,t),
\end{eqnarray}
where $\vec{u}$ denotes a suitable flow velocity of nucleons. Now, we can employ the closure relation,
\begin{eqnarray} \label{eq18}
\sum _{a}\Phi _{a}^{*} (\vec{r}_{1} ,t)\Phi _{a} (\vec{r}_{2} -\vec{u}\tau ,t) =\delta (\vec{r}_{1} -\vec{r}_{2} +\vec{u}\tau ),
\end{eqnarray}
where the summation $a$ runs over the complete set of states originating from target or projectile, and the closure relation is valid for each set of the
spin-isospin degrees of freedom. The flow velocity $\vec{u}(\vec{R},T)$ may depend on the mean position $\vec{R}=(\vec{r}_{1} +\vec{r}_{2} )/2$ and
the mean time $T=(t+t')/2$. Employing the closure relation in the first term of the right hand side of Eq.\,(\ref{eq16}), we find
\begin{eqnarray} \label{eq19}
\sum _{a\in P,h\in T}\!\!\!\!\!A_{ah}^{\alpha } (t)A_{ah}^{*\alpha } (t')&=&\sum _{h\in T}\int d^{3} r_{1}  d^{3} r_{2} \delta (\vec{r}_{1} -\vec{r}_{2} +\vec{u}_{h} \tau )\nonumber\\
&&\qquad\;\times W_{h}^{\alpha } (\vec{r}_{1} ,t)W_{h}^{*\alpha } (\vec{r}_{2} ,t').
\end{eqnarray}
In this manner the complete set of single-particle states is eliminated and the calculation of the expression is greatly simplified. In fact, in
order to calculate this expression, we only need the hole states originating from target which are provided by the TDHF description.  Rather than
the mean flow velocity, we take local flow velocity $\vec{u}_{h} (\vec{R},T)$ of each hole state across the window for each term in the summation. The
local flow velocity of each wave-function is specified by the usual expression of the current density divided by the particle density as given in Eq.(\ref{eqa6}) in Appendix\,\ref{sec:app}. The quantity $W_{h}^{\alpha } (\vec{r}_{1} ,t)$ becomes
\begin{eqnarray} \label{eq20}
\!\!\!\!W_{h}^{\alpha } (\vec{r}_{1} ,t)=\frac{\hbar }{m} g(x_{1} )\left[\nabla _{1} \Phi _{h}^{\alpha } (\vec{r}_{1} ,t)-\frac{x_{1} }{2\kappa ^{2} } \Phi _{h}^{\alpha } (\vec{r}_{1} ,t)\right],
\end{eqnarray}
and $W_{h}^{*\alpha } (\vec{r}_{2} ,t')$ is given by a similar expression.  A detailed analysis of Eq.\,(\ref{eq19}) under certain approximation is
presented in Appendix\,\ref{sec:app}.  The result of this analysis as given by Eq.\,(\ref{eq16}) is,
\begin{equation} \label{eq21}
\sum _{a\in P,h\in T}\!\!\!\!A_{ah}^{\alpha } (t)A_{ah}^{*\alpha } (t')=G(\tau )\int d^{3} r \tilde{g}(x)
J_{X,\alpha }^{T} (\vec{r},t-\tau /2).
\end{equation}
Here $J_{x,\alpha }^{T} (\vec{r},t-\tau /2)$ represents the sum of the magnitude of the current densities due to wave functions originating from
target and it is given by Eq.\,(\ref{eqa17}). The quantity $G(\tau )$ is the average value of the memory kernels $G_{h} (\tau )$ given by Eq.\,(\ref{eqa18}). It is
possible to carry out a similar analysis in the second term in the right side of Eq.\,(\ref{eq13}) which yields,
\begin{equation} \label{eq22}
\sum _{a\in T,h\in P}\!\!\!\!A_{ah}^{\alpha } (t)A_{ah}^{*\alpha } (t')=G(\tau )\int d^{3} r \tilde{g}(x)
J_{x,\alpha }^{P} (\vec{r},t-\tau /2).
\end{equation}
In a similar manner, $J_{x,\alpha }^{P} (\vec{r},t-\tau /2)$ is determined by the sum of the magnitude of the current densities due wave functions
originating from projectile, and it is given by an equation similar to Eq.\,(\ref{eqa17}). In Eq.\,(\ref{eq21}) and Eq.\,(\ref{eq22}) we use lower case $\vec{r}$
instead of capital letter. As a result, the quantal expressions of the proton and neutron diffusion coefficients are given by,
\begin{eqnarray} \label{eq23}
D_{\alpha \alpha } (t)&=&\int _{0}^{t}d\tau  G(\tau )\int d^{3} r \tilde{g}(x)\left[J_{x,\alpha }^{T} (\vec{r},t-\tau /2)\right.\nonumber\\
&&\qquad\qquad\qquad\qquad\quad\left.+J_{x,\alpha }^{P} (\vec{r},t-\tau /2)\right]\nonumber\\
&&-\int _{0}^{t}d\tau\text{Re}\left[\sum _{h'\in P,h\in T}\!\!\!\!A_{h'h}^{\alpha } (t)A_{h'h}^{*\alpha } (t-\tau )\right.\nonumber\\
&&\qquad\qquad\left.+\!\!\sum _{h'\in T,h\in P}\!\!\!\!A_{h'h}^{\alpha } (t)A_{h'h}^{*\alpha } (t-\tau ) \right].
\end{eqnarray}
To our knowledge, such a quantal expression for the nucleon diffusion coefficient in heavy-ion collisions is given in the literature for the first
time from a microscopic basis. There is a close analogy between the quantal expression and the classical diffusion coefficient in a random walk
problem\,\cite{randrup1979,gardiner1991,weiss1999}. The first line in the quantal expression gives the sum of the nucleon currents from the target-like fragment to
the projectile-like fragment, which is integrated over the memory. This is analogous to the random walk problem, in which the diffusion coefficient is
given by the sum of the rate for the forward and backward steps.  The second line in the quantal diffusion expression stands for the Pauli blocking
effect in nucleon transfer mechanism, which does not have a classical counterpart. It is important to note that the quantal diffusion coefficients
are entirely determined in terms of the occupied single-particle wave functions obtained from the TDHF solutions.

In the calculations carried out for the present study, we find that the average nucleon flow speed across the window between the colliding nuclei is
around $u_{x} \approx 0.05$c. Using the expression $\tau {}_{0} =\kappa /|u_{x} |$  given below Eq.\,(\ref{eqa18}), with a dispersion $\kappa =1.0$ fm, we find
the memory time to be around $\tau _{0} \approx 20$ fm/c. In the nuclear one-body dissipation mechanism, it is possible to estimate the memory time in
terms of a typical nuclear radius and the Fermi speed as $\tau _{0} \approx R/v_{F} $. If we take $R\approx 5.0$ fm and $v_{F} \approx 0.2$c, we find
the same order of magnitude for the memory time, $\tau _{0} \approx 25$ fm/c. Since it is much shorter than a typical interaction time of collisions at sub-barrier energies,
$\tau _{0}\ll 400$ fm/c, we find that the memory effect is not very effective in nucleon exchange mechanism. We neglect the memory effect in the first
line of the diffusion coefficient. The time integration of the memory kernel alone becomes,
\begin{eqnarray} \label{eq24}
\tilde{G}(t)=\int _{0}^{t}d\tau  G(\tau )&=&\int _{0}^{t}d\tau  \frac{1}{\sqrt{4\pi }\tau _{0} } e^{-(\tau /2\tau _{0} )^{2} }\nonumber\\
&=&\frac{1}{2} \text{erf}(t/2\tau _{0} ).
\end{eqnarray}
Because of the same reason, memory effect is not very effective in the Pauli blocking terms as well, however in the calculations we keep the memory
integrals in these terms.

\section{Results of Calculations}
\label{sec:4}
Employing the quantal diffusion mechanism described in the previous section, we investigate nucleon exchange mechanism in the central collisions of
$^{28}$O + $^{28}$O, $^{40}$Ca + $^{40}$Ca, $^{48}$Ca + $^{48}$Ca, and $^{56}$Ni + $^{56}$Ni systems at bombarding energies slightly below the fusion
barriers. 
Calculations were done in a three-dimensional
Cartesian geometry with no symmetry assumptions\,\cite{umar2006c} and using the
Skyrme SLy4 interaction\,\cite{chabanat1998a}.
The three-dimensional Poisson equation for the Coulomb potential
is solved by using Fast-Fourier Transform techniques
and the Slater approximation is used for the Coulomb exchange term.
The box size used for all the calculations
was chosen to be $60\times 30\times 30$~fm$^3$, with a mesh spacing of
$1.0$~fm in all directions. These values provide very accurate
results due to the employment of sophisticated discretization
techniques\,\cite{umar1991a}.

\begin{figure}[!hpt]
\includegraphics[width=8cm]{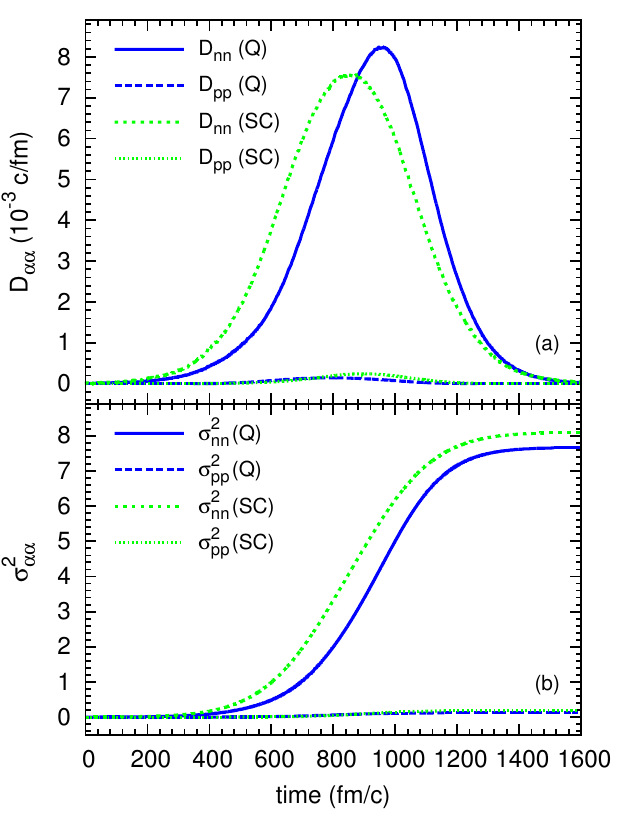}
\caption{(color online) Quantal and semi-classical neutron and proton diffusion coefficients (a) and corresponding variances (b) in central collisions of ${}^{28}
\text{O}+{}^{28} \text{O}$ at $E_{c.m.} =8.7$ MeV.\label{fig1}}
\end{figure}

\begin{figure}[!hpt]
	\includegraphics[width=8cm]{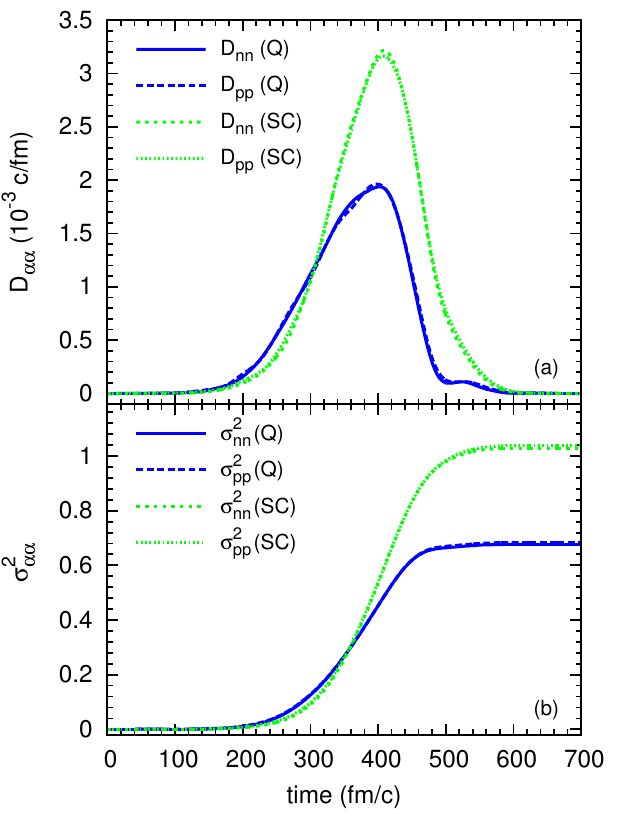}
	\caption{(color online) Quantal and semi-classical neutron and proton diffusion coefficients (a) and corresponding variances (b) in central collisions of ${}^{40}
		\text{Ca}+{}^{40} \text{Ca}$ at $E_{c.m.} =52.7$ MeV.\label{fig2}}
\end{figure}

\begin{figure}[!hpt]
	\includegraphics[width=8cm]{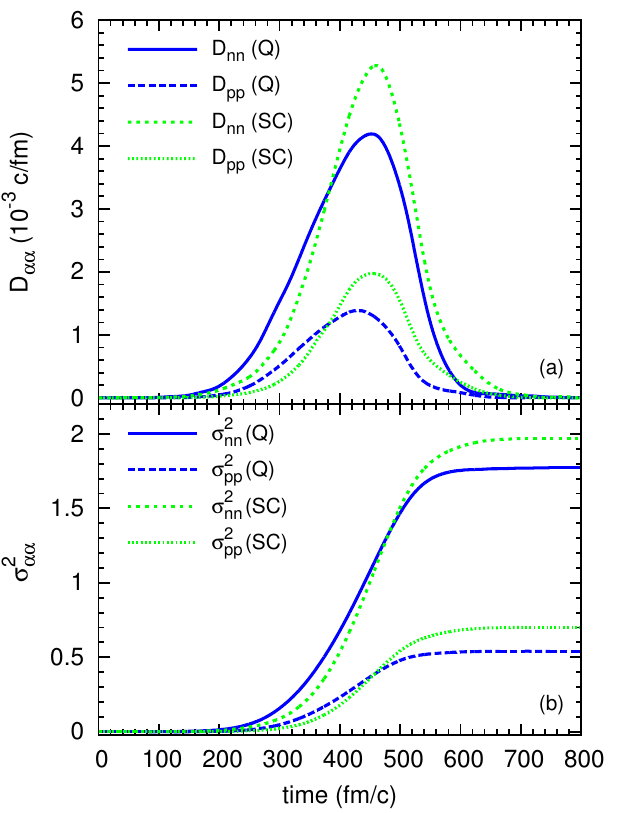}
	\caption{(color online) Quantal and semi-classical neutron and proton diffusion coefficients (a) and corresponding variances (b) in central collisions of ${}^{48}
		\text{Ca}+{}^{48} \text{Ca}$ at $E_{c.m.} =50.7$ MeV.\label{fig3}}
\end{figure}

\begin{figure}[!hpt]
	\includegraphics[width=8cm]{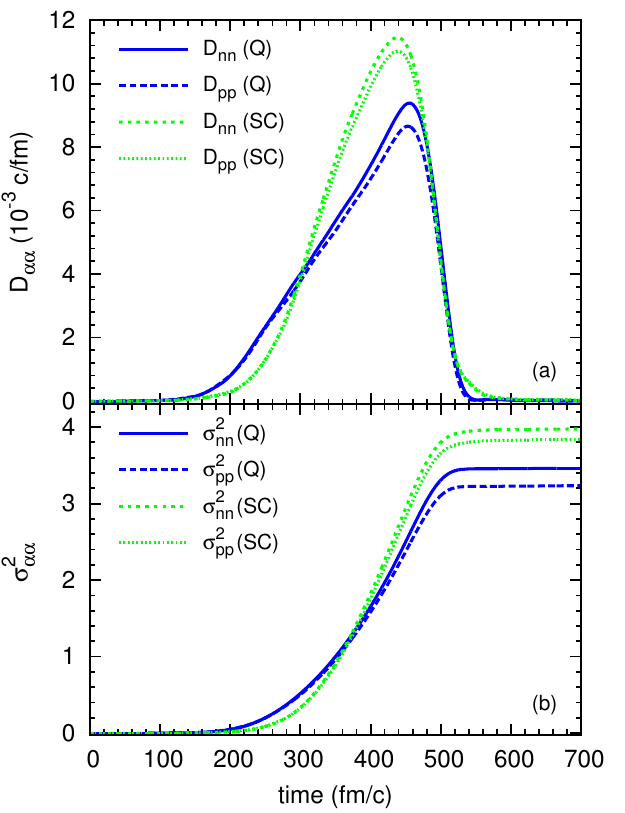}
	\caption{(color online) Quantal and semi-classical neutron and proton diffusion coefficients (a) and corresponding variances (b) in central collisions of ${}^{56}
		\text{Ni}+{}^{56} \text{Ni}$ at $E_{c.m.} =100$ MeV.\label{fig4}}
\end{figure}

In Table\,\ref{tab:1}, we present the fusion barriers and the bombarding energies at which the calculations are carried out for these systems.
During the reactions, colliding ions stick together with a visible neck for some time, and separate without forming a compound nucleus. Because of the
symmetry, the mean values of proton and neutron numbers of the separated fragments remain equal to their initial values. However, proton and neutron
numbers of the outgoing fragments have distributions around their mean values with variances determined by diffusion coefficients as,
\begin{eqnarray} \label{eq25}
\sigma _{\alpha \alpha }^{2} (t)=2\int _{0}^{t}dt' D_{\alpha \alpha } (t').
\end{eqnarray}
In a number of previous studies, we carried out calculations by employing the semi-classical approximation of the
diffusion coefficient. We can obtain the semi-classical approximation of the diffusion coefficient by taking the Wigner transform of Eq.\,(\ref{eq23}).
\begin{table}[!ht]
	\centering
	\caption{\label{tab:1}The fusion barriers and bombarding energies of the systems. The energies are given in MeV units.}
	\begin{tabular}{|c|c|c|}
		\hline
		System & Fusion Barrier & Bombarding Energy\\
		\hline
		$^{28}$O + $^{28}$O & 8.8 & 8.7 \\
		\hline
		$^{40}$Ca + $^{40}$Ca & 53.2 & 52.7 \\
		\hline
		$^{48}$Ca + $^{48}$Ca & 51.0 & 50.7 \\
		\hline
		$^{56}$Ni + $^{56}$Ni & 100.7 & 100.0 \\
		\hline
	\end{tabular}
\end{table}
In this manner, it is possible to express diffusion coefficient in terms of the phase-space distribution functions associated with single-particle
wave functions originating from target and projectile nuclei. The semi-classical diffusion coefficients have similar form that is familiar from the
nucleon exchange transport model\,\cite{randrup1979}. In order to avoid negative regions of the phase-space functions an averaging procedure is carried out
as outlined in\,\cite{yilmaz2011,yilmaz2014,ayik2015a}. Such an averaging procedure is particularly important for an accurate description of the Pauli blocking
effects. In our presentation, we compare the quantal diffusion coefficients and the quantal variances for neutron and proton distributions of the
outgoing fragments with their semi-classical approximation corresponding to same reactions. The result of calculations of diffusion coefficients and
variances for the systems investigated are presented in Figs.\,\ref{fig1}-\ref{fig4}.
The upper panels (a) of the figures show diffusion coefficients and the lower panels (b) illustrate variances as a function of times for the systems investigated. Solid lines and long dashed-lines indicate the quantal results for neutrons and protons, respectively. Similarly, short dashed-lines and dotted-lines show the semi-classical results for neutrons and protons, respectively.
The observed differences between the quantal and the semi-classical calculations originate from three different sources. The quantal calculations
naturally include shell effects while in the semi-classical calculations the shell effects are washed out. In the mean-field description of TDHF approach
the collective motion is treated in near classical approximation, but the single-particle motion is treated in a fully quantal manner. Therefore, in
the quantal calculations, the barrier penetration of nucleons across the window is fully accounted for. On the other hand, in the semi-classical
calculations only those nucleons above the barrier are allowed to cross the window. particularly at low energies the barrier penetration in nucleon
transfer can make a big difference for both protons and neutrons. In Table\,\ref{tab:2}, we list the asymptotic values of the proton and neutron variance for
system investigated. We refer to the contribution for the part of the variances arising from the first line in the diffusion coefficient in Eq.(\ref{eq23}) as direct term, and refer to the term due to the second line as the blocking term.
\begin{table}[ht]
\centering
\caption{\label{tab:2}Effect of Pauli blocking on fragment neutron and proton variances. The bombarding energies of all systems are given in Table\,\ref{tab:1}. Abbreviations Q and SC
stand for quantal and semi-classical, respectively. PB stands for Pauli blocking and $\Delta$ is the difference between the variances with and without
Pauli blocking.}
\begin{tabular}{|c|c|c|c|c|c|c|c|}
\cline{3-8}
\multicolumn{2}{c|}{\multirow{2}{*}{}}&\multicolumn{3}{c|}{$\sigma_{nn}^2(t\rightarrow\infty)$}&\multicolumn{3}{c|}{$\sigma_{pp}^2(t\rightarrow\infty)$}\\
\cline{3-8}
\multicolumn{2}{c|}{}& with PB &no PB&$\Delta_{nn}$& with PB &no PB&$\Delta_{pp}$\\
\hline
\multicolumn{1}{|c|}{\multirow{2}{*}{$^{28}$O}}&Q&7.66&9.57&-1.91&0.12&0.28&-0.16\\
\cline{2-8}
&SC&8.10&8.72&-0.62&0.18&0.15&0.03\\
\hline
\multicolumn{1}{|c|}{\multirow{2}{*}{$^{40}$Ca}}&Q&0.67&1.51&-0.84&0.68&1.52&-0.84\\
\cline{2-8}
&SC&1.03&1.11&-0.08&1.04&1.12&-0.08\\
\hline
\multicolumn{1}{|c|}{\multirow{2}{*}{$^{48}$Ca}}&Q&1.77&3.17&-1.4&0.54&1.21&-0.67\\
\cline{2-8}
&SC&1.97&2.35&-0.38&0.70&0.72&-0.02\\
\hline
\multicolumn{1}{|c|}{\multirow{2}{*}{$^{56}$Ni}}&Q&3.46&5.44&-1.98&3.23&5.25&-2.02\\
\cline{2-8}
&SC&3.98&4.20&-0.22&3.84&4.09&-0.25\\
\hline
\end{tabular}
\end{table}
We observe by comparing the second column for neutrons and the second column for protons in Table\,\ref{tab:2} that the
direct contributions in the variances in the quantal calculations are larger than the semi-classical results
(mainly as a result of the barrier penetration). The third important
difference between the quantal and the semi-classical results arises from the Pauli blocking terms in the diffusion coefficient. In the quantal
calculations the Pauli blocking terms are calculated exactly. On the other hand, the Pauli blocking effects in the semi-classical limit are treated in
an approximate manner. By comparing the third column for neutrons and the third column for protons in Table\,\ref{tab:2}, we notice
 large differences in the
magnitude of the Pauli blocking terms between the quantal and the semi-classical calculations. In fact,for some situations the averaging procedure to
eliminate the negative regions of the phase-space functions may not work very well, consequently sign of Pauli blocking terms can become positive
rather than negative. Because of these different effects, the asymptotic values of the proton and neutron variances for the quantal calculations may be
smaller or larger that the result of the semi-classical calculations. Even the small differences in the variances can make a large effect on the
production of rare neutron rich isotope by the diffusion mechanism.  It is important to note that the quantal diffusion calculations are not only more
accurate, but also the fact that the quantal calculations take less numerical effort than their semi-classical counterparts.

\section{Conclusions}
\label{sec:conc}
In the standard mean-field approach, the collective motion is treated semi-classically. The SMF approach improves the standard
description by incorporating thermal and quantal fluctuations in the initial states. In this manner, the SMF approach provides an approximate
description of the quantal fluctuation dynamics of collective motion at low energies where collisional dissipation is not very effective. Under
certain circumstances, the fluctuation dynamics can be approximately described in terms of transport coefficients associated with the collective
variables.  In this work, we consider central collisions of symmetric nuclei below fusion barrier and study nucleon exchange mechanism in the SMF
approach. Since binary structure is maintained during the collision, it is possible to determine macroscopic variables by a geometric projection
procedure. The SMF approach, gives rise to a Langevin description for evolution of macroscopic variables. In this work, we consider nucleon exchange
mechanism in the central collisions of symmetric nuclei and extract quantal expression for the diffusion coefficients of proton and neutron
exchanges.  We carry out calculations of proton and neutron variances in the central collisions of ${}^{28} \text{O}+{}^{28} \text{O}$, ${}^{40}
\text{Ca}+{}^{40} \text{Ca}$, ${}^{48} \text{Ca}+{}^{48} \text{Ca}$,  and ${}^{56} \text{Ni}+{}^{56} \text{Ni}$ systems at bombarding energies
slightly below the fusion barriers, and compare the quantal results with the corresponding semi-classical calculations. There are
important differences between the quantal and the semi-classical calculations due to mainly three different mechanisms. First of all the quantal
calculations involve shell effects, while the shell effects are smoothed out in the semi-classical calculations. The barrier penetration of
protons and neutron during the transfer across the window are properly taken into account in the quantal description. In the semi-classical
calculations, nucleon transfers are totally blocked below the barrier of the mean-field potential.  More importantly, in the quantal description, the
Pauli blocking effects in the transfer mechanism is taken into account exactly, while in the semi-classical calculations it is taken into account in an
approximate manner. The quantal calculations provide more accurate description of the diffusion coefficients, but surprisingly they require less
numerical effort than the semi-classical calculations.

\begin{acknowledgments}
S.A. gratefully acknowledges the IPN-Orsay and the Middle East Technical University for warm hospitality extended to him during his visits. S.A. also
gratefully acknowledges useful discussions with D. Lacroix. This work is supported in part by US DOE Grant No. DE-SC0015513, and in part by US DOE
Grant No. DE-SC0013847.
\end{acknowledgments}

\appendix
\section{Analysis of the Closure Relation}
\label{sec:app}
We re-write Eq.\,(\ref{eq16}) as,
\begin{eqnarray}
\label{eqa1}
\!\!\!\!\!\!\!\sum _{a\in P,h\in T}\!\!\!\!A_{ah}^{\alpha } (t)A_{ah}^{*\alpha } (t')&=&\sum _{h\in T}\int d^{3} R d^{3} r\delta (\vec{r}+\vec{u}_{h} \tau )\nonumber\\
&&\qquad\times W_{h}^{\alpha } (\vec{r}_{1} ,t)W_{h}^{*\alpha } (\vec{r}_{2} ,t'),
\end{eqnarray}
where we introduce the coordinate transformation,
\begin{eqnarray}
 \vec{R}=\left(\vec{r}_{1} +\vec{r}_{2}\right)/2\;,\quad\vec{r}=\vec{r}_{1}-\vec{r}_{2},
\end{eqnarray}
and its reverse as
\begin{eqnarray}
\label{eqa3}
\vec{r}_{1}=\vec{R}+\vec{r}/2,\quad \vec{r}_{2} =\vec{R}-\vec{r}/2.
\end{eqnarray}
For clarity, we present quantities $W_{h}^{\alpha } (\vec{r}_{1} ,t)$ and $W_{h}^{*\alpha } (\vec{r}_{2} ,t')$  here again,
\begin{eqnarray}
 W_{h}^{\alpha } (\vec{r}_{1} ,t')&=&\frac{\hbar }{m} g\left(X+\frac{x}{2}\right)\left[\nabla_{X}i _{h}^{\alpha }\left(\vec{R}+\frac{\vec{r}}{2},t'\right)\right.\nonumber\\
&&\qquad\left.-\frac{X+x/2}{2\kappa ^{2}}\Phi _{h}^{\alpha }\left(\vec{R}+\frac{\vec{r}}{2} ,t'\right)\right]
\end{eqnarray}
and
\begin{eqnarray}
 W_{h}^{*\alpha } (\vec{r}_{2} ,t')&=&\frac{\hbar }{m} g\left(X-\frac{x}{2}\right)\left[\nabla_{X} \Phi _{h}^{*\alpha }\left(\vec{R}-\frac{\vec{r}}{2},t'\right)\right.\nonumber\\
&&\qquad\left.-\frac{X-x/2}{2\kappa ^{2}}\Phi _{h}^{*\alpha }\left(\vec{R}-\frac{\vec{r}}{2} ,t'\right)\right]
\end{eqnarray}
The local flow velocity of the wave function $\Phi _{h}^{\alpha } (\vec{R},T)$ is calculated in the standard manner,
\begin{eqnarray}\label{eqa6}
\vec{u}_{h}^{\alpha } (\vec{R},T)&=&\frac{\hbar }{m} \frac{1}{|\Phi _{h}^{\alpha } (\vec{R},T)|^{2} }\nonumber\\
&&\times\text{Im}\left[\Phi _{h}^{*\alpha } (\vec{R},T)\vec{\nabla} \Phi _{h}^{\alpha } (\vec{R},T)\right],
\end{eqnarray}
with $T=(t+t')/2=t-\tau /2$. Because of the delta function in the integrand of Eq.\,(\ref{eqa1}), we make the substitution  $\vec{r}=-\vec{u}_{h}^{\alpha } (\vec{R},T)\tau $ in the wave functions and introduce
the backward diabatic shift to obtain,
\begin{eqnarray}
 \Phi _{h}^{\alpha } (\vec{R}+\vec{r}/2,t)&=&\Phi _{h}^{\alpha } (\vec{R}-\vec{u}_{h}^{\alpha } \tau /2,t)\nonumber\\
&\approx& \Phi _{h}^{\alpha } (\vec{R},T),
\end{eqnarray}
and
\begin{eqnarray}
 \Phi _{h}^{\alpha } (\vec{R}-\vec{r}/2,t')&=&\Phi _{h}^{\alpha } (\vec{R}+\vec{u}_{h}^{\alpha } \tau /2,t')\nonumber\\
&\approx& \Phi _{h}^{\alpha } (\vec{R},T).
\end{eqnarray}
After making this substitution, Eq.\,(\ref{eqa1}) becomes,
\begin{eqnarray}
\label{eqa9}
&&\frac{m^2}{\hbar^2}\!\!\!\sum _{a\in P,h\in T}\!\!\!\!A_{ah}^{\alpha } (t)A_{ah}^{*\alpha } (t')\nonumber\\
 &=&\sum _{h\in T}\int d^{3} R  \tilde{g}(X)\frac{G_{h} (\tau )}{|u_{X}^{h} (\vec{R},T)|} \left[ |\nabla _{X} \Phi _{h}^{\alpha } (\vec{R},T)|^{2}\right. \nonumber\\
&&- \frac{X}{2\kappa ^{2} }  \nabla_{X} \left(|\Phi _{h}^{\alpha } (\vec{R},T)|^{2} \right)\nonumber\\
&&+ \left.\frac{X^{2} -(u_{X}^{h} \tau /2)^{2} }{4\kappa ^{4} } |\Phi _{h}^{\alpha } (\vec{R},T)|^{2}\right].
\end{eqnarray}
In this expression $\tilde{g}(X)$ is sharp as Gaussian smoothing function centered on the window with a dispersion $\kappa =0.5$ fm,
\begin{eqnarray}
\tilde{g}(X)=\frac{1}{\sqrt{\pi } \kappa } \exp [-(X/\kappa )^{2} ],
\end{eqnarray}
and $G_{h} (\tau )$ indicates the memory kernel,
\begin{eqnarray}
G_{h} (\tau )=\frac{1}{\sqrt{4\pi } } \frac{1}{\tau _{o}^{h} } \exp [-(\tau /2\tau _{o}^{h} )^{2} ],
\end{eqnarray}
with the memory time $\tau _{0}^{h} =\kappa /|u_{X}^{h} |$.  Due to the fact that $\tilde{g}(X)$  is centered at $X=0$, the second term in Eq.\,(\ref{eqa9}) is
nearly zero.  In the third term, after carrying out an average over the memory, the factor in the middle becomes,
\begin{eqnarray}
X^{2} -(u_{x}^{h} \tau /2)^{2} \to X^{2} -(\kappa /2)^{2}.
\end{eqnarray}
Since Gaussian $\tilde{g}(X)$ is sharply peaked around $X=0$ with a variance $(\kappa /2)^{2} $, the third terms in Eq.\,(\ref{eqa9}) is expected to be very
small, as well.  Neglecting the second and third terms, Eq.\,(\ref{eqa9}) becomes,
\begin{eqnarray}
\label{eqa13}
&&\frac{m^2}{\hbar^2}\sum _{a\in P,h\in T}\!\!\!\!A_{ah}^{\alpha } (t)A_{ah}^{*\alpha } (t')\nonumber\\
&=&\sum _{h\in T}\int d^{3} R  \tilde{g}(X)\frac{G_{h} (\tau )}{|u_{X}^{h} (\vec{R},T)|} |\nabla _{X} \Phi _{h}^{\alpha } (\vec{R},T)|^{2}.
\end{eqnarray}
Furthermore, it is useful to express the wave functions in terms of its magnitude and its phase as\,\cite{gottfried1966},
\begin{eqnarray}
\Phi _{h}^{\alpha } (\vec{R},T)=|\Phi _{h}^{\alpha } (\vec{R},T)|\exp \left[iQ_{h}^{\alpha } (\vec{R},T)\right].
\end{eqnarray}
The phase factor $Q_{h}^{\alpha } (\vec{R},T)$ behaves as the velocity potential of the flow velocity of the wave. Using the definition given by Eq.\,(\ref{eqa6}), 
we observe that the flow velocity is given by $\vec{u}_{h}^{\alpha } (\vec{R},T)=(\hbar /m)\vec{\nabla }Q_{h}^{\alpha } (\vec{R},T)$.
In the vicinity of the window, in the perpendicular direction, the phase varies faster than the magnitude of the wave function. Neglecting the small
variation of the magnitude $|\Phi _{h} (\vec{R},T)|$, we can express the gradient of the wave function in Eq.\,(\ref{eqa13}) as,
\begin{eqnarray}
\nabla _{X} \Phi _{h}^{\alpha } (\vec{R},T)\approx i\Phi _{h}^{\alpha } (\vec{R},T)\nabla _{X} Q_{h}^{\alpha } (\vec{R},T).
\end{eqnarray}
As a result, Eq.\,(\ref{eqa1}) becomes,
\begin{eqnarray}
\label{eqa16}
&&\sum _{a\in P,h\in T}\!\!\!\!A_{ah}^{\alpha } (t)A_{ah}^{*\alpha } (t')\nonumber\\
&=&G(\tau )\int d^{3} R \tilde{g}(X)J_{X,\alpha }^{T} (\vec{R},t-\tau /2).
\end{eqnarray}
Here, the quantity $ J_{X,\alpha }^{T} (\vec{R},t-\tau /2) $ in the integrand  represents the sum of the magnitude of the current densities due to wave functions originating from target,
\begin{eqnarray}\label{eqa17}
&&J_{X,\alpha }^{T} (\vec{R},t-\tau /2)\nonumber\\
&&\!\!\!=\frac{\hbar }{m} \sum _{h\in T}|\text{Im}\left[\Phi _{h}^{*} (\vec{R},t-\tau /2)\nabla _{X} \Phi _{h} (\vec{R},t-\tau /2)\right]|.
\end{eqnarray}
In obtaining Eq.\,(\ref{eqa16}), we introduced a further approximation by replacing the individual memory kernels $G_{h} (\tau )$ by its average value taken over the hole states,
\begin{eqnarray}
\label{eqa18}
 G(\tau )=\frac{1}{\sqrt{4\pi } } \frac{1}{\tau _{0} } \exp [-(\tau /2\tau _{0} )^{2} ] ,
\end{eqnarray}
with the memory time determined by the average  speed $u_{X} $ as $\tau _{0} =\kappa /|u_{X} |$.

\bibliography{VU_bibtex_master}

\end{document}